\documentclass[prd,preprint]{revtex4}

\usepackage{fancyhdr,lastpage,graphicx}
\usepackage{amsmath}
\usepackage[shortlabels]{enumitem}
\usepackage[paper=letterpaper,
            marginparwidth=1.2in,     
            marginparsep=.05in,       
            margin=1in,               
            includemp]{geometry}
\begin{document}
 \newcommand{\bq}{\begin{equation}}
 \newcommand{\eq}{\end{equation}}
 \newcommand{\bqn}{\begin{eqnarray}}
 \newcommand{\eqn}{\end{eqnarray}}
 \newcommand{\bqs}{\begin{equation}\begin{split}}
 \newcommand{\eqs}{\end{split}\end{equation}}
 \newcommand{\nb}{\nonumber}
 \newcommand{\lb}{\label}
 \newcommand{\p}{\partial}

\title{Experimental Test of the Final State Hypothesis}

\begin{abstract}
The black hole final state projection model, also known as the Horowitz-Maldacena\cite{mald,unit} model has garnished new interest due to the current debate over black hole firewalls. The nonlinear quantum mechanics of post-selection preserves information and avoids the AMPS\cite{sam,amps} argument by relaxing monogamy of entanglement. While these are promising features there are also potentially observable predictions to be made.
\end{abstract}

\author{Michael J Devin}
\maketitle

One proposed resolution of the black hole information paradox\cite{hawk} is known as the black hole final state of Horowitz-Maldacena conjecture.\cite{mald} The simple description of which is that singularities such as those in black holes, are areas in which all quantum fields take on unique boundary values. It can also be regarded as stating that singularities have zero entropy. Since black hole singularities for ordinary non-extreme cases are thought to be space-like, then the boundary condition on these entities is similar to an initial state for an ordinary system, only here the surface is in the proper future of observers, thus the name final state. The situation classically might appear overdetermined, but quantum mechanically there is some wiggle room found in the physics of post-selected ensembles.\cite{ahr1}\\

In post selection, branches of the wave function that violate the selection condition, such as having an incorrect value approaching the singularity, are removed, and the remaining state renormalized to conserve probability. If the condition defines a linear subspace, then the process can be expressed as a projection operation, hence the name final state projection. This deformation of regular quantum mechanics can be used to simulate odd constraints on systems. One relevant example is using post selection to enforce periodicity of a qubit, thus simulating a time machine.\cite{me,seth1,me2}  The system considered is essentially the same as the quantum teleportation protocol\cite{tele1,vaidm}, with two exceptions. First that classical reference qubit required to decode the teleported bit is selected to a known constant or final state , thus removing the need for communicating it's value through the classical channel. Second, without the classical channel there is no reason that the 'destination bit' needs to be in the forward light-cone of the 'originating' bit. If it is instead in the 'past' light-cone, then we have a simulated time machine. Though limited by decoherence effects, it is sufficient to model many circuits and phenomena to gain a better understanding of causality in this deformation of quantum mechanics.\\

One of the important scenarios to examine in this simulation is the grandfather paradox\cite{seth1}. Post-selection relies on a nonzero normalization factor which in certain cases can be shown to vanish. A more general prescription of approximate post-selection can avoid this, but with its' own cost. Like any good paradox, there is something to learn. While the discarded normalization factor may not affect the relative probabilities of measurements after the chosen in-state, it can also be seen as a relative weight or prior probability of the prescribed initial state.\cite{me,me2} This is a natural extension of the idea of paradox censorship for possible grandfather paradoxes in simulated time travel through post-selection. The prior probability of states leading to a paradox goes to zero just as the ensemble size does in a simulation.\cite{seth1} Because causality is more ambiguous in post-selected systems it is useful to consider all probabilities as conditional, based upon the assumption of reference states that may be before or after a measurement.\\

Early attacks on post-selection as a mechanism for saving information in black holes revolved around these normally anathema violations of causality, as well as the ensuing grandfather paradox.\cite{presk} It is clear from the simulated grandfather paradox scenario, that interactions like those suggested by Preskill in \cite{presk} during collapse will be suppressed to the extent that they lead to a vanishing of the normalization factor. Other efforts to shore up the behavior to avoid paradoxes are redundant, as the nature of post-selection is to avoid paradoxes. This is not to say there are no problems. The skewing of prior probabilities that prevents paradox during collapse back-propagates onto the pre-collapse matter and Hawking radiation, as well as any normal incoming radiation.\\

Post-selected quantum mechanics is nonlinear, and as such is subject to certain effects common to other nonlinear extensions of quantum mechanics, namely state duplication, over-entanglement, and loss of causality. However these effects are exactly what is counted on to avert the loss of information to the singularity. A few have proposed experiments to detect violations of quantum mechanics through the change of entanglement non-locally.\cite{me,yuri1} However these approaches suffer from three problems. First, there is the lack of easy access to evaporating black holes to manipulate potential Hawking radiation. Second, we have relatively limited coherence length and time scales for systems we might possibly use to reach a black hole, compared to the distance and time required for any experiment. Lastly, the large environmental noise factor associated with any system passing through an accretion disc should swamp any interference signal we hope to find.\\

The mechanism of teleporting qubits is the center of the HM model. In-falling Hawking radiation\cite{hawk75} is forced to cancel the fields coming from the collapsing star, by post-selection. Then the outgoing HR which is still entangled fully with it's incoming part, will similarly correlate to the in-falling star. One of the main problems with this simple model is that if we consider the HR to be produced as some small distance from the horizon, either at plankian distances for a stretched horizon or from the further angular potential at $3m$, then it's incoming part must catch up to the star in order to cancel it. (See figures 1-3)\\

Consider the internal black hole metric, and Penrose diagram. Despite the name, the singularity is still an extended object, having a spatial length stretching from the formation to the evaporation of the black hole. A classical horizon problem emerges for the in-falling system. Particles that originate at late times and fall into the hole, cannot reach the same 'part' of the singularity as the original star.\cite{bousso} Backwards light cones near the singularity pinch up, and regions separated by the now spatial time co-ordinate lose causal contact with each other before ending. Either some mechanism must allow the state hitting the singularity to propagate along it's length, despite the vanishing local speed of light in the transverse direction, or nearly all of the incoming HR pairs must be created shortly after the star crosses it's trapping surface. If the radiation must be emitted from a plank length away from the horizon, then the entire black hole must evaporate very rapidly, perhaps on the order of the mixing time. Due to conservation of energy, it would appear that all of the collapsing star simply piled up and was reflected off of the horizon just as the black hole formed.(see figures 4,5) \\

This could be thought of as a realization of the stretched horizon\cite{suss}, but there is no reason to believe it would go unnoticed by observers falling in with the collapsing star. Post-selection may partially suppress interaction between the in-falling star and its time reversed image. Time reversed qubits that are decohered by interaction with the in-falling matter will generally reduce the post-selection normalization constant by a factor exponential in the number of interactions, as the system becomes more like a classical time machine. Post-selection may also help spread some of the in-falling star out along the singularity, increasing the black hole's lifetime, but it is difficult to say by how much. This is because the trade off between forcing all of the pairs to be emitted from the low temperature of the early black hole radiation, and the 'entropy cost' of transporting the qubits to the relatively higher temperature late times. The relative weights of the two phenomena depend on the behavior of the normalization constant for each diagram, as well as the details of very high energy interaction near the singularity.  Several competing statistical effects are at work. Seen from another perspective we can think of HM as equivalently imposing a boundary condition very near the horizon.(see figure 6) Since a surface just inside the horizon acts as a Cauchy surface for the interior, we can imagine backward evolving the boundary condition from near the singularity outward. The horizon acts then as a very rough mirror, since we assume a smooth boundary near the singularity would normally be surrounded by a very high entropy mixing era just prior to it from the point of view of the collapsing matter. The 'proper temporal ordering' gives a hard firewall at the horizon.\cite{sam,amps}\\

This needs not be the case. The horizon behavior can be mitigated by carefully choosing the singularity boundary condition.\cite{unit} Since the boundary condition acts as a one time pad encryption in the middle of the teleportation process, it is trivially true that we can fine tune the singularity boundary condition to eliminate the firewall. However, this boundary condition should properly be a constant of the theory, not tunable on a case by case basis. If we choose a condition to allow the black hole to have a long lifetime, that late time radiation is still not capable of receiving the teleported states from the collapsing star. Instead a large part of the 'time reversed scrambled image' of the star will 'go out' as ordinary radiation. Specifically a wavefront of incoming energy that 'compliments' the star.(see figure 7) Studying the behavior of the normalization factor for various collapse scenarios will show the prior probability preference for a complimentary incoming mode is a generic property of boundary conditions that lead to weaker firewalls. If we want to avoid a firewall not all of the time reversed modes can reflect into the HR.\\

A few possible astronomical consequences arise from these considerations. For some stringent boundary conditions we might expect no black hole as such to even form\cite{hawk06}, but for the horizon to reflect all fields. Black hole states in that model would be deselected, and post-selection would create statistical forces that prevent further collapse. These forces could significantly affect stellar dynamics, and lead to more energetic novas for stars above the Chandrasekhar mass. Physics is strongly different in an otherwise low curvature region.(see figure 8) Complementary radiation may have a less dramatic effect. The profile of such un-emissions would be short and strong, similar to currently studied gamma ray bursts. They could reasonably be expected to have evaded detection for much the same reason gamma bursters did for so long. The difference is that they would not be directly observable by normal detectors, since the events would be like time reversed gamma burst detections. They would largely fail to penetrate the atmosphere for the same reason, and different non-absorbing detectors would be needed. Recoil events similar to those of WIMPS might give them away, as well as other indirect means of measurement. The correlation of several such events over multiple detectors in both time and direction would be a very specific signature of the theory.\\
\begin{figure}[h]
\centering
\includegraphics[scale=.4]{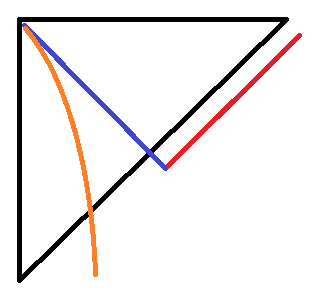}
\caption{A typical collapsing body (orange) meets up with incoming Hawking radiation(blue) near one end of the singularity. The outgoing radiation (red) must be produced early enough to reach the matter.}
\end{figure}
\begin{figure}[h]
\centering
\includegraphics[scale=.35]{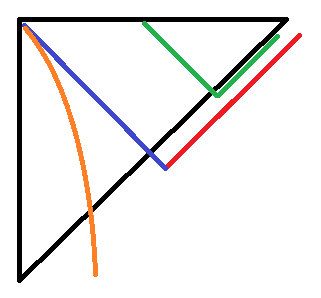}
\caption{Late time radiation may be supressed or decorrelated from the original star.  Later pairs hit another region of the singularity(green), and unless cancelled by late in-falling matter or another Hawking pair, may be prevented by the boundary condition.}
\end{figure}
\begin{figure}[h]
\centering
\includegraphics[scale=.4]{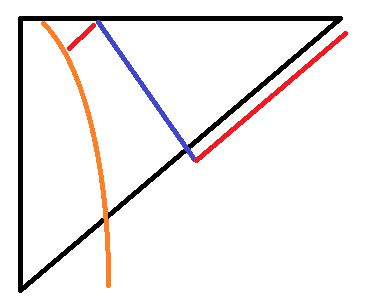}
\caption{For large black holes tidal forces and pressures may only do so much to spread the fields out along the singularity if they do not become strong until well after crossing the horizon.}
\end{figure}
\begin{figure}[h]
\centering
\includegraphics[scale=.4]{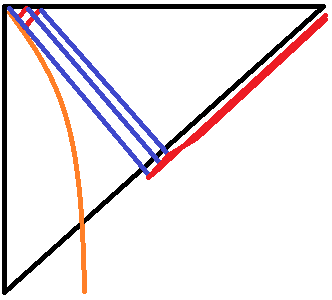}
\caption{A black hole becomes a hollow fuzzball in a short time. All of the information is transferred to the 'stretched horizon' region(red), through the time reversed image(blue). Post-selection may supress the interaction between the star and it's image, giving rise to a sort of complementarity.}
\end{figure}
\begin{figure}[h]
\centering
\includegraphics[scale=.4]{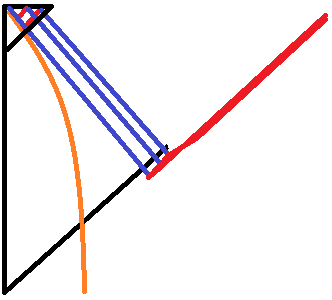}
\caption{With nearly all the star's mass just outside the horizon, There is arguably no longer a black hole, but only an apparent horizon and a strange wall of negative energy particles(blue).}
\end{figure}
\begin{figure}[h]
\centering
\includegraphics[scale=.4]{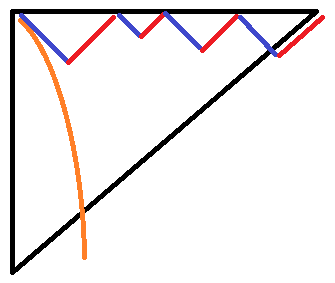}
\caption{A qubit bouncing laterally along the singularity to be emmitted at late times. A large number of particle 'coincidences' may be required to emit radiation at a steady rate, but this is balanced by the alternative, a similarly improbable burst of early particles. The exponential decay of the normalization factor for such behavior is similar to the exponential decay of horizon perturbations.}
\end{figure}
\begin{figure}[h]
\centering
\includegraphics[scale=.3]{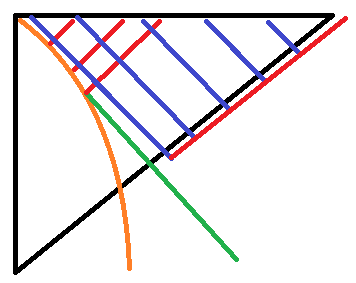}
\caption{Incoming rays help cancel and diffuse the star along the singularity, leading to slower decay. Late radiation may still be uncorrelated, as some information is 'lost' into the 'complimentary in states'(green).}
\end{figure}
\begin{figure}[h]
\centering
\includegraphics[scale=.4]{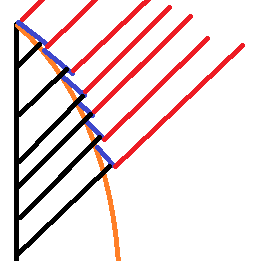}
\caption{Another possibility, the rapid disintegration of the star as it collapses, preventing any black hole from forming.}
\end{figure}

\end{document}